\documentclass{main}

\usepackage{textcomp}
\usepackage{fancyhdr}
\usepackage{lastpage}
\usepackage{subcaption}

\def\be{\begin{equation}}
\def\ee{\end{equation}}
\def\bea{\begin{eqnarray}}
\def\eea{\end{eqnarray}}

\newcommand{\Photo}{\includegraphics[height=35mm]{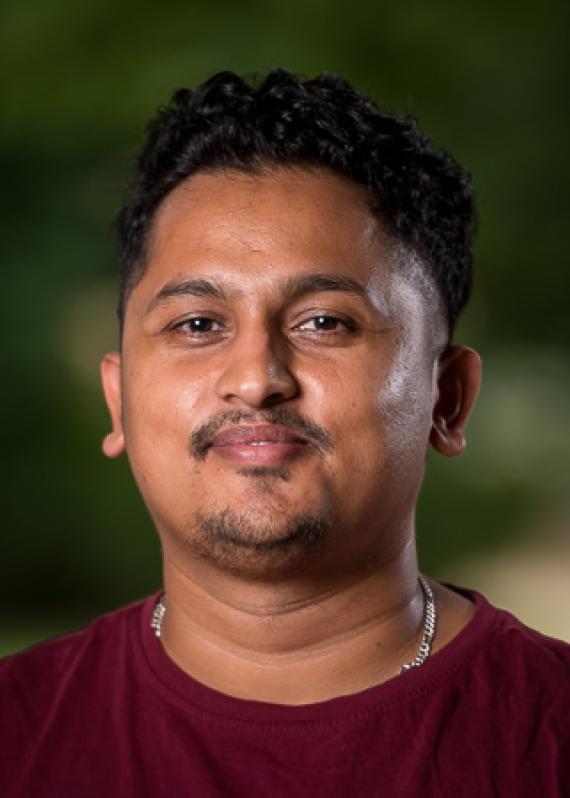}}

\fancypagestyle{firstpagefooter}{%
  \fancyfoot[L]{  \textcopyright \footnotesize This work is an open access article under a Creative Commons Attribution 4.0 International License (https://creativecommons.org/licenses/by/4.0/).}
  \fancyfoot[C]{}  
   
}

\begin{document}

\thispagestyle{firstpagefooter}
\title{\Large First Study of Exclusive Production of Multi-Hadron Final States in Ultraperipheral Collisions }

\author{\underline{Amrit Gautam for the ALICE Collaboration}\footnote{Speaker, email: amritgautam@ku.edu} }

\address{
Department of Physics and Astronomy, The University of Kansas, Lawrence, KS, 66045, USA
}

\maketitle\abstracts{
The study of exclusive photoproduction of multi-hadron final states in ultra-peripheral collisions (UPCs) provides a unique avenue to explore quantum chromodynamics (QCD) and the nature of resonances emerging from gluonic interactions. The ALICE Collaboration has recently performed measurements of exclusive four-pion photoproduction using Run 2 data, favoring the presence of two resonances. However, the underlying nature of these resonances remains poorly understood. The enhanced capabilities of the ALICE detector during Run 3 open new opportunities for the study of multi-pion final states, including four and six-pion systems, and processes involving charmonium decays into four-hadron final states. These studies provide an invaluable way to probe the dynamics of highly dense gluonic matter and the interplay between resonant and non-resonant contributions in hadronic systems.
In this paper, we outline the ALICE program focused on these analyses.
}

\footnotesize DOI: \url{https://doi.org/xx.yyyyy/nnnnnnnn}


\section{Introduction}
\begin{figure}
    \centering
    \includegraphics[width=0.2\linewidth]{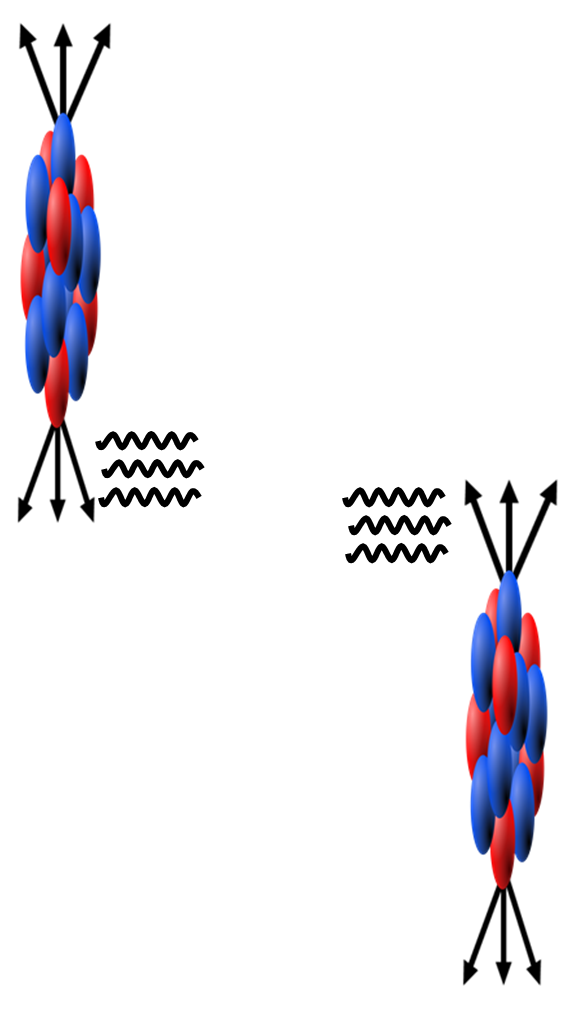}
    \caption{Lorentz contracted nuclei moving near the speed of light. The interaction is dominated by the exchange of a virtual photon when \it{b} $>$ 2\it{R}}
    \label{fig:upc_nuclei}
\end{figure}

\begin{figure}
    \centering
    \includegraphics[width=0.2\linewidth]{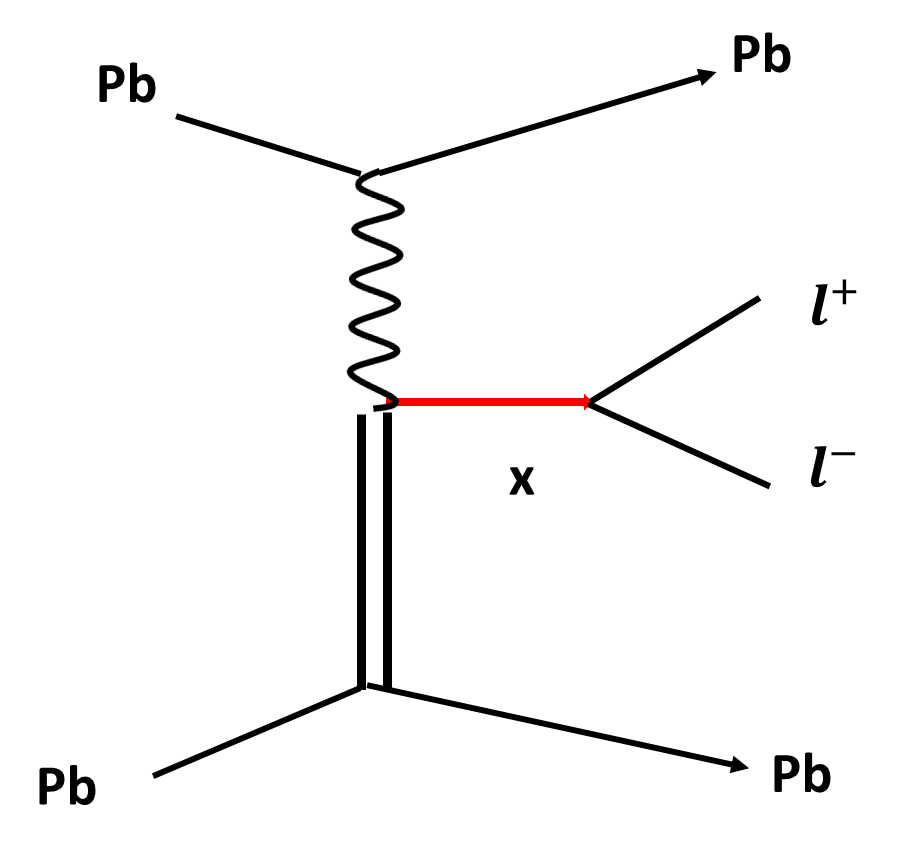}
    \caption{Exclusive production of a vector meson in UPCs, which then decays further.}
    \label{fig:upcfyn}
\end{figure}

The CERN Large Hadron Collider (LHC) is a particle accelerator capable of accelerating protons and ions near the speed of light to unprecedented energies. These particles can collide at different points using powerful magnets, and the result of the collision can be observed using state-of-the-art detectors. A Large Ion Collider Experiment (ALICE) is located at point 2 of the LHC and is composed of multiple sub-detectors. ALICE is particularly interested in colliding heavy ions, nominally Pb. In particular, when the impact parameter is greater than twice the radius of the ions, the interaction is mediated by quasi-real photons, as shown in Fig.\ref{fig:upc_nuclei} and \ref{fig:upcfyn}. The collisions involving electromagnetic interactions are known as ultra-peripheral collisions (UPCs)~\cite{phyupcLHC,Contreras:2015dqa,Klein:2019qfb}. UPCs most frequently produce particles called vector mesons. The vector mesons are particles made out of a quark and an anti-quark pair, having spin 1. The vector mesons are the result of the interaction between the quasi-real photon from the electromagnetic field of the source nuclei with the target. At ALICE, during Run 1 and Run 2, vector mesons such as $J/\psi$, $\rho^{0}$ have been extensively studied ~\cite{ALICE:2012yye,ALICE:2023polarization,ALICE:jpsi_psi}. The nuclei typically remain intact during UPCs, and the resulting number of particles are relatively low. These vector mesons are sensitive to gluon dynamics in the target, and its internal nuclear structure. UPCs have proven to be a useful tool to study as part of the CERN LHC heavy-ion physics program. Recent advancements in detector technology have led to the upgrade of the ALICE detector for Run 3~\cite{ALICE:2023upgrade}. The upgrade enabled a streaming readout, achieving higher statistics. 
In this document, recent results from the Run 2 period, focusing on the azimuthal correlation of coherent $\rho^0$ and the first measurement of excited $\rho^0$, will be discussed. Moreover, the physics performance studies of multi-prong final states with the new data from Run 3 will be presented. 

\section{Azimuthal correlations of coherent \texorpdfstring{$\rho^0$}{rho} mesons}
\label{sec:prod}
During a UPC interaction, photons from one nucleus can interact with the other nucleus as a target. In a coherently photoproduced $\rho^0$, photons interact coherently with the entire nucleus. However, among two nuclei, any nucleus can act as a source and target~\cite{interference_vect}. Hence, there are two possibilities. Since the separation is at the femtometer scale, the quantum effect becomes dominant and we can observe the quantum mechanical interference between such two possibilities. Experimentally, this effect can be observed as anisotropy in the angular distribution of $\pi^+$ and $\pi^-$, which are decay products of $\rho^0$. 

The ALICE Collaboration has recently made a detailed analysis~\cite{ALICE:2024ife}. The data were collected during the Run 2 period with colliding Pb ions at a center-of-mass energy of $\sqrt{s_{\rm{NN}}}=5.02$ TeV , and a dedicated UPC trigger. As UPC events are characterized by having a few tracks and almost no activity in the forward direction, detectors close to the beam pipe, such as V0 and AD detectors, are used as a veto. During Run 2, this was done both for the online and offline selections, while during Run 3, such selections are done only offline. The trigger in Run 2 for UPC events required no activity in forward detectors, within the time window of beam-beam interactions.
Additionally, the trigger also had a topological requirement using the Silicon Pixel Detector (SPD). At least two SPD tracklest with an opening angle greater than 153 degrees were required. This was because coherently produced $\rho^0$ mesons have relatively low transverse momentum and decay into two back-to-back pions. The pion tracks were reconstructed using information from the Inner Tracking System (ITS) and the Time Projection Chamber (TPC). 

The strength of this interference effect varies with the distance between colliding nuclei. This distance can be controlled in data using an independent process called Coulomb excitation ~\cite{PhysRev.96.237_coulomb}.  During Coulomb excitation, multiple photons can be exchanged independently, and can sometimes excite the target nuclei. The excited nuclei can then dissociate by producing neutrons along the beam. These neutrons can be measured using Zero Degree Calorimeters (ZDC). The impact parameter can be estimated using the number of detected neutrons. A large number of neutrons detected in the ZDC would indicate a small impact parameter, and from zero to a few neutrons would indicate a larger impact parameter. Experimentally, neutron production was divided into 3 classes: 0n0n, meaning no neutrons on both sides of the collision (large impact parameter); 0nXn, meaning only one neutron on one side of the collision (medium impact parameter), and XnXn, where there is at least one neutron on both sides (large impact parameter). The nOOn Monte Carlo generator is used for modeling the neutron emission~\cite{Broz:2019kpl}

\begin{figure}[ht]
    \centering
    \includegraphics[width=0.7\linewidth]{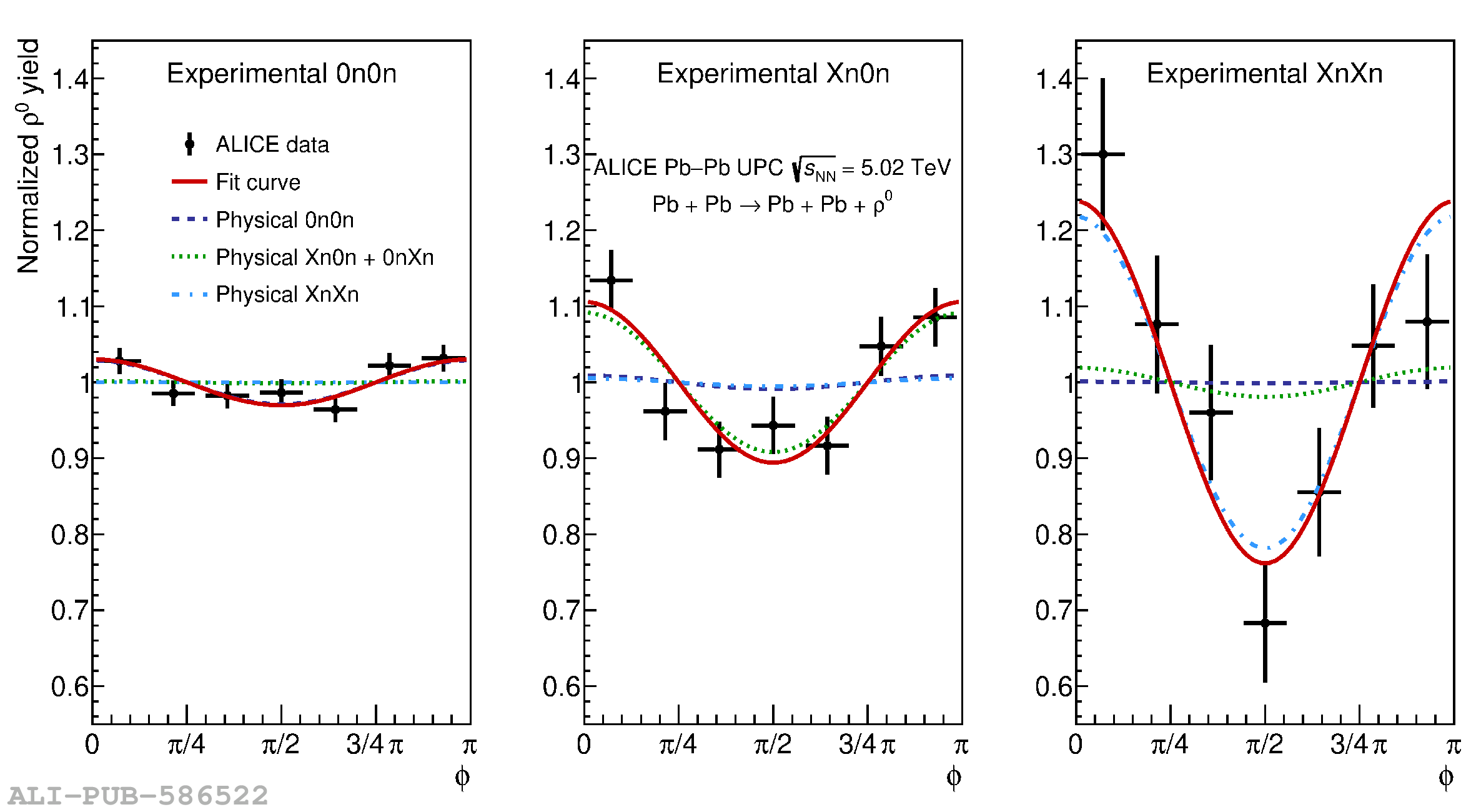}
    \caption{The measured $\varphi$ distribution for three different neutron classes used to extract the angular anisotropy. The leftmost plot represents the 0n0n class (large impact parameter); the middle plot represents 0nXn (medium impact parameter), and the rightmost plot represents XnXn (small impact parameter). The red curve shows the simultaneous fit performed to extract the anisotropy as a component of $\cos~2\varphi$.}
    \label{fig:azimuth}
\end{figure}

Figure~\ref{fig:azimuth} shows the distribution of azimuthal angle $\varphi$. Here $\varphi$ is defined as the angle between the transverse component of $p_+(\pi_1 +\pi_2)$ and $p_-(\pi_1 -\pi_2)$. The $\pi_1$ and $\pi_2$ are chosen randomly from $\pi^+$ and $\pi^-$. This definition removes the $cos~\varphi$ component, so any observed anisotropy appears as a component of $cos~2\varphi$ term. In Fig.~\ref{fig:azimuth}, the leftmost plot shows the $\varphi$ distribution in the 0n0n class, corresponding to large impact parameters. Very small effect is observed in the angular distribution, as expected. The interference effect becomes more dominant for the smallest impact parameter in the XnXn class. 

\begin{figure}
    \centering
    \includegraphics[width=0.6\linewidth]{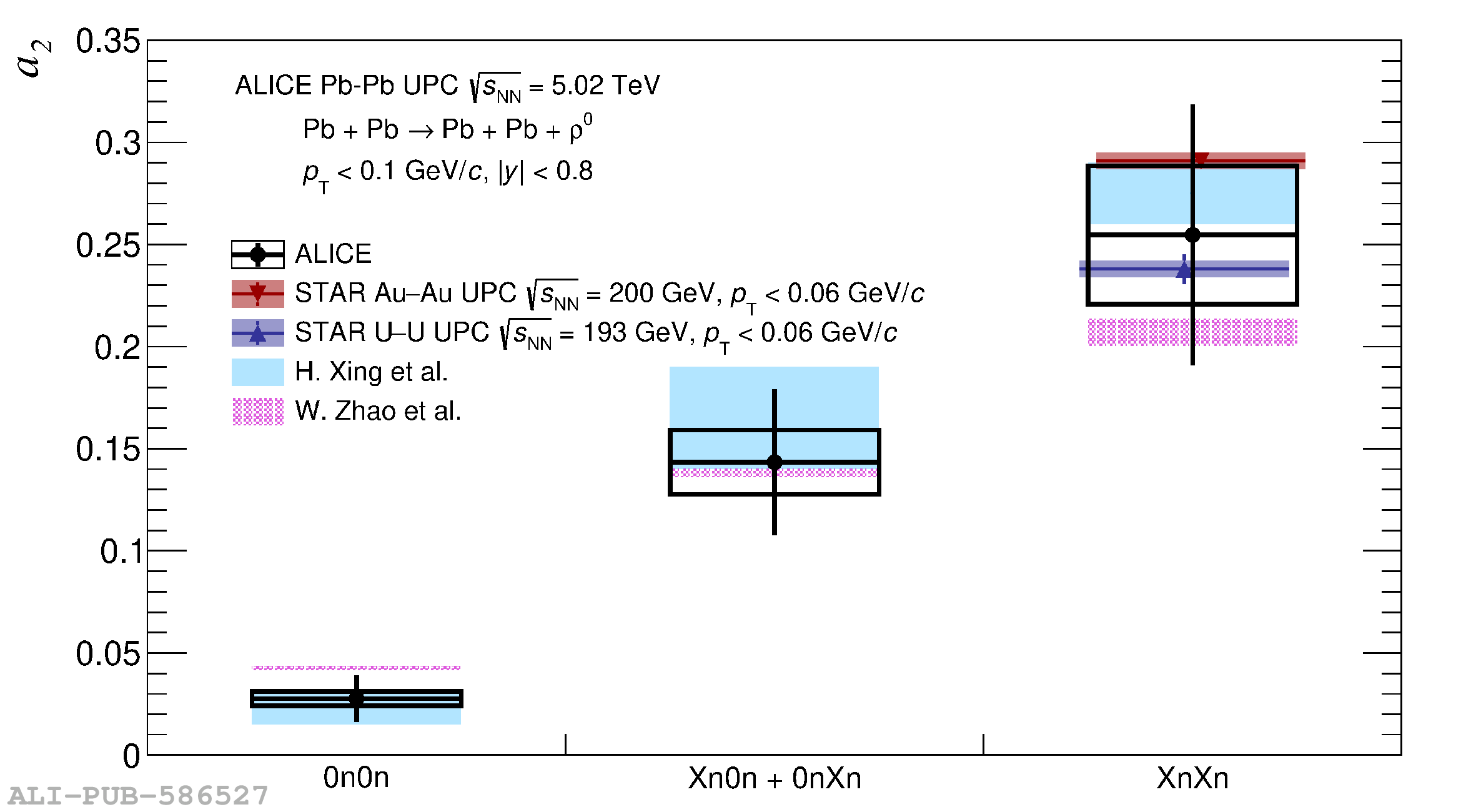}
    \caption{Measured coefficient $\it{cos~}$$ 2\varphi$ compared with multiple models and previous measurements.}
    \label{fig:aniso}
\end{figure}

This study further quantifies this effect as the anisotropy for each neutron class using the amplitude of $cos~2\varphi$ extracted from the fit of Fig.\ref{fig:azimuth}. The \autoref{fig:aniso} compares the extracted amplitude with two models \cite{Xing2020}~\cite{WZHa}. The result is also compared with previous measurements by STAR~\cite{star_tomo} for Au-Au and U-U collisions at center-of-mass energies of $\sqrt{s_{\rm{NN}}} =  200$ and $\sqrt{s_{\rm{NN}}}=193$~GeV, respectively. While the ALICE measurement performed for this study agrees with STAR for XnXn class, this measurement further extends the measurement to the 0nXn and 0n0n classes. This measurement also agrees with the models within uncertainties. The ALICE measurement is the first study of the impact parameter dependence of the $\varphi$ modulation for coherently photoproduced $\rho^0$ mesons. An order of magnitude increase in the modulation from 0n0n to XnXn, from high to low impact parameter, is observed. 

\section{First measurement of excited \texorpdfstring{$\rho$}{rho} in the four-pion decay channel in UPC at LHC}
\label{sec:section_four_pi}
\begin{figure}
    \centering
    \includegraphics[width=0.6\linewidth]{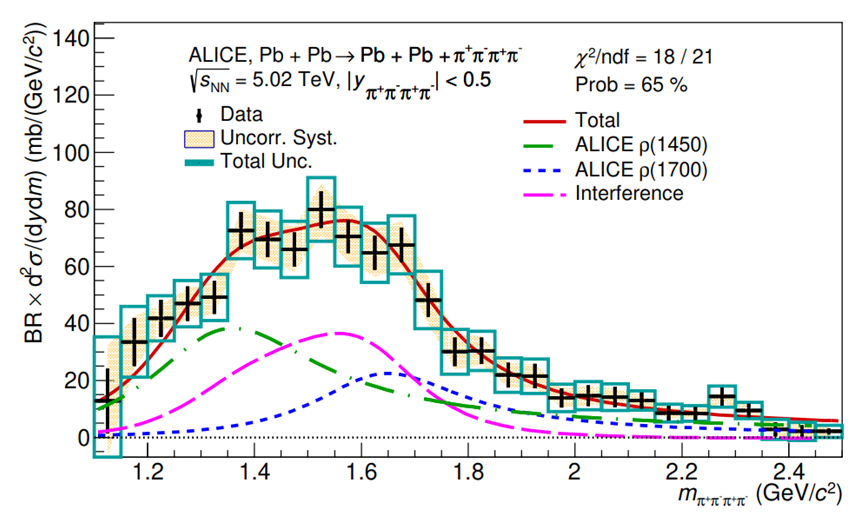}
    \caption{Invariant mass of four pions in UPC Pb--Pb collisions at LHC at $\sqrt{s_{\rm{NN}}} = 5.02$ TeV.}
    \label{fig:fourpi_mass}
\end{figure}

The coherent production of $\rho^0 (770)$ in UPCs has been extensively studied at RHIC and LHC (see references in~\cite{ALICE:2024kjy}). The Particle Data Group (PDG) discusses the presence of the excited state of the $\rho^0$ vector mesons called $\rho'$~\cite{ptep_pdg}. The PDG also lists multiple resonances of excited $\rho^0$ vector mesons: $\rho(1450)$ and $\rho(1700)$. The experimental measurements by the OMEGA spectrometer ~\cite{WA91:1994tzu_4pi_omega} and STAR~\cite{STAR:2009giy} observed the presence of these excited states in exclusive production of four-pion events. However, recent preliminary results from H1\cite{H1prelim2018} in electron-proton collisions suggest that one single broad resonance around 1500 MeV/$\it{c}^{2}$ may describe the data. It is important to study this process at the CERN LHC energies to clarify the nature of these resonances. For this study~\cite{ALICE:4pi}, the data collected in ALICE during Run 2 were utilized. The trigger and data sample were the same as for the azimuthal correlation analysis described above. However, instead of the two pion final state, four pions with net zero charge were selected. The \autoref{fig:fourpi_mass} shows the invariant mass of the four coherently produced pions after efficiency and acceptance correction. 

\begin{figure}[hh]
    \centering
    \includegraphics[width=0.5\linewidth]{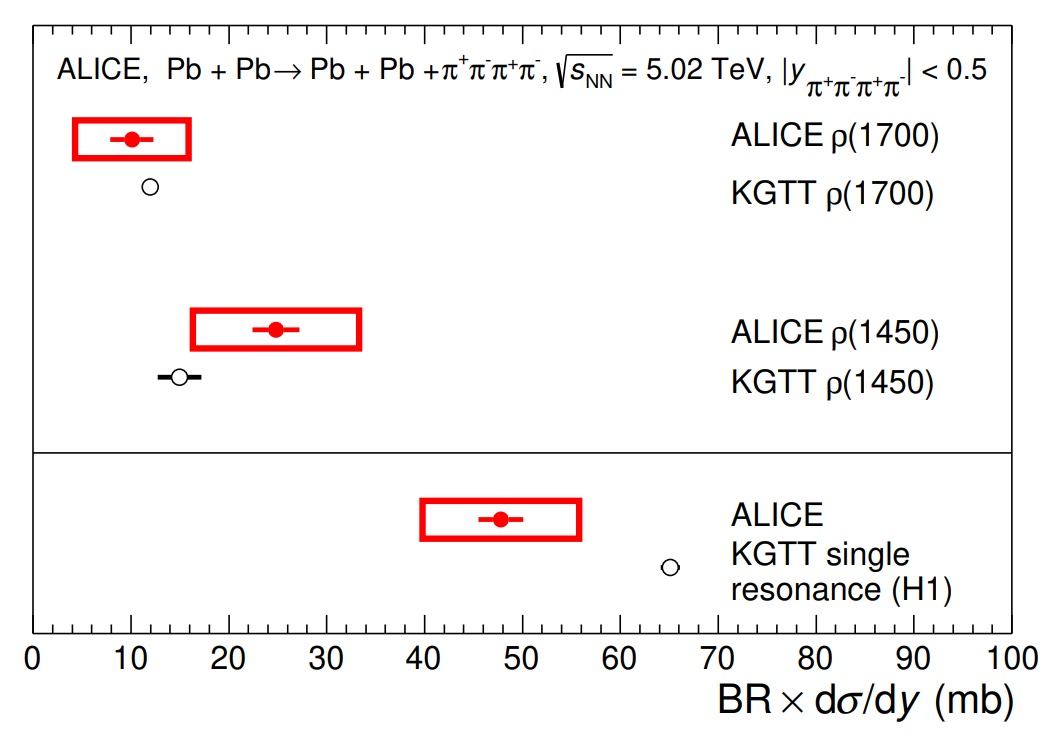}
    \caption{Comparison of the invariant mass fit using two-resonance hypotheses and a single-resonance hypothesis with the KGTT theoretical model.} \label{fig:four_pi_hypo_th}
\end{figure}

The invariant mass distribution shows a peak from $\approx$1 GeV/$c^2$ to $\approx$2 GeV/$c^2$. This broad peak was fitted with two resonance hypotheses and includes the interference between these two resonances. The fit of the single-resonance hypothesis and double-resonance hypothesis is compared with the theoretical model KGTT~\cite{Klusek-Gawenda:2020gwa}. The comparison between the KGTT model is shown in Fig.~\ref{fig:four_pi_hypo_th}. From the comparison, one can conclude the presence of two resonances: $\rho(1450)$ and $\rho(1700)$. This result provides a high-precision measurement for this process. At the same time, more data from the LHC Run 3 period will be needed for more systematic studies ~\cite{Klusek-Gawenda:2020gwa,Devi:2025ftf}.

\section{Run 3 studies}
\begin{figure}[h]
    \centering
    \begin{subfigure}{0.5\linewidth}
        \centering
        \includegraphics[width=\linewidth]{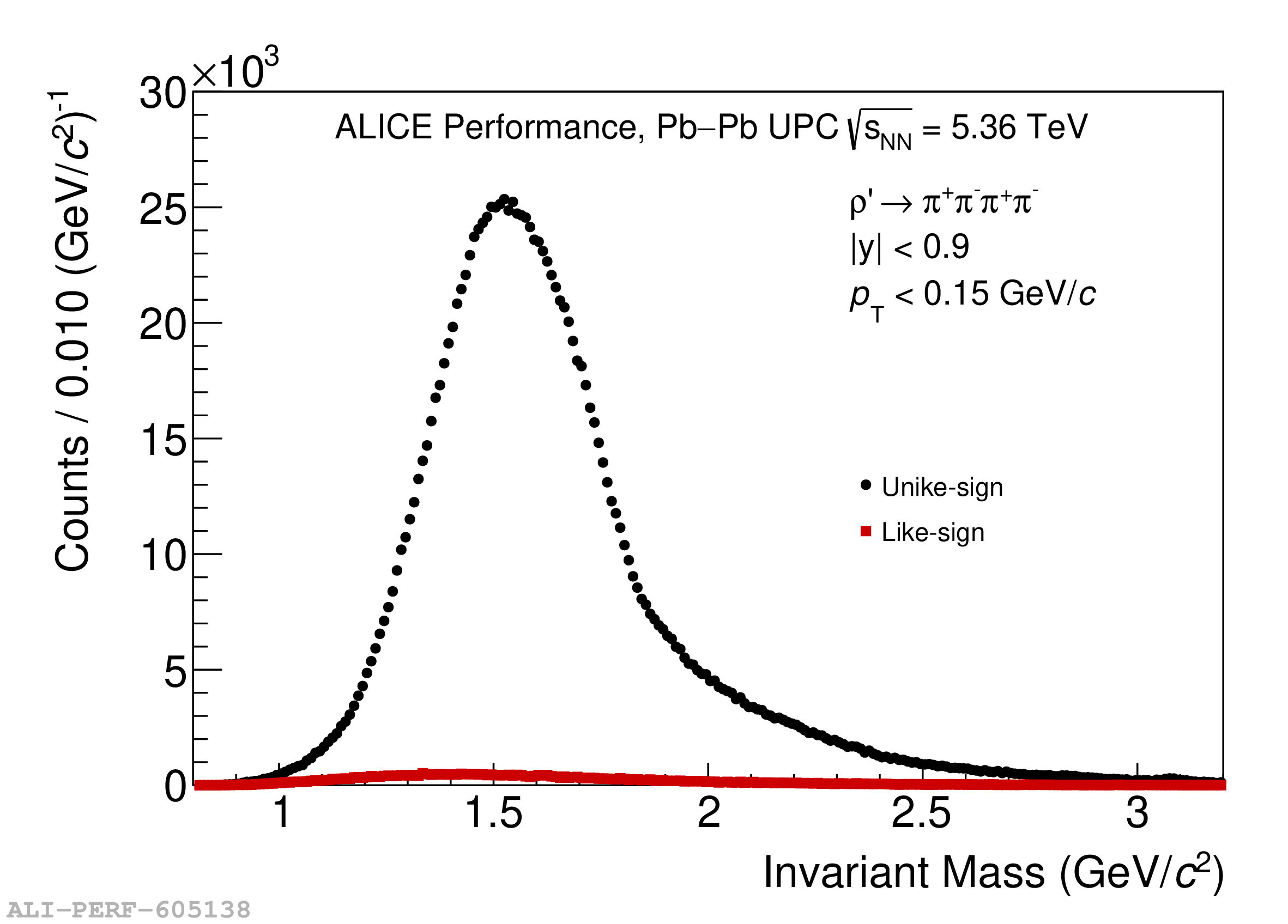}
        \caption{}
        \label{fig:rho_prime_run3}
    \end{subfigure}
    \hfill
    \begin{subfigure}{0.48\linewidth}
        \centering
        \includegraphics[width=\linewidth]{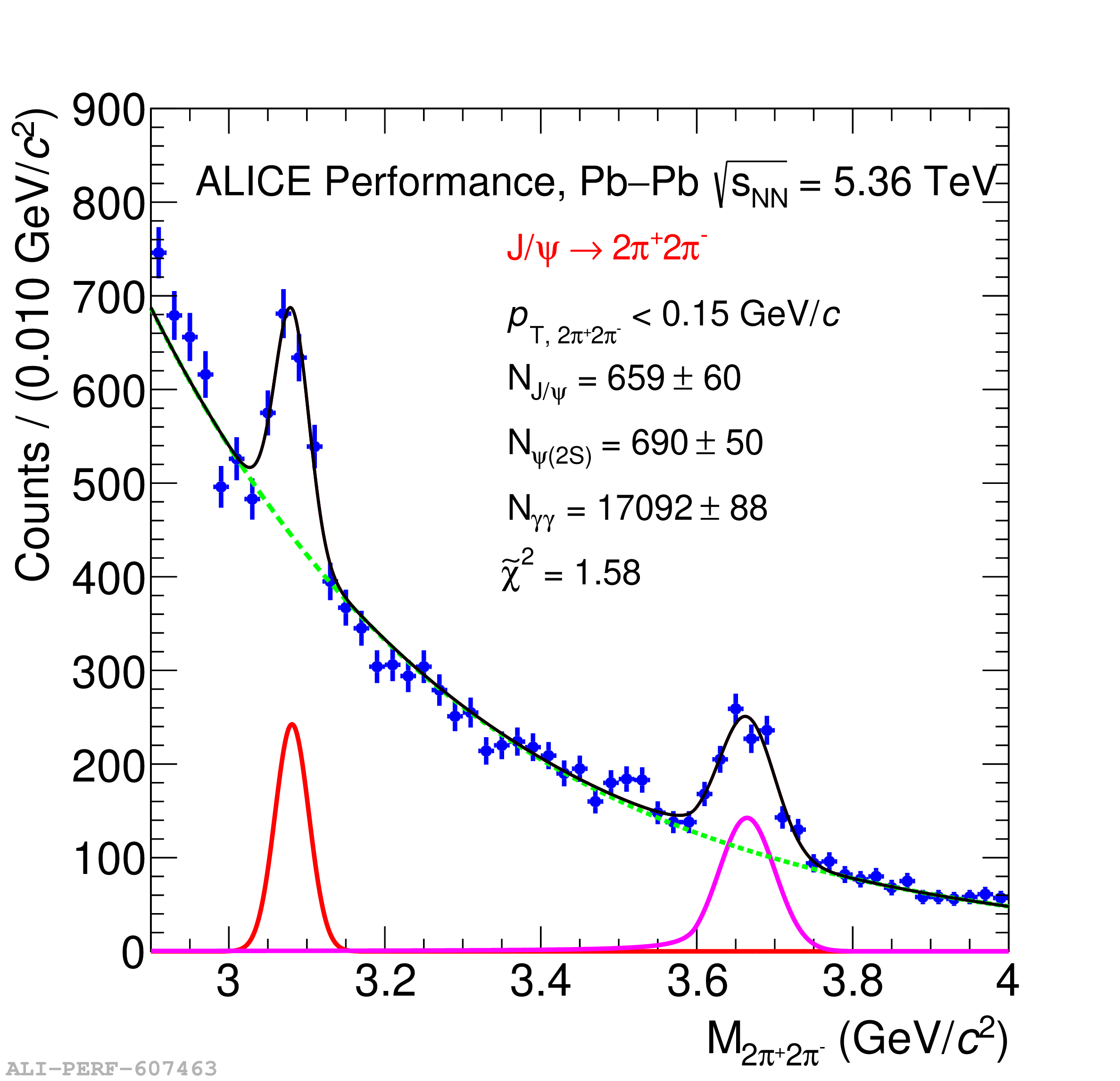}
        \caption{}
        \label{fig:jptofourpi_run3}
    \end{subfigure}
    \caption{Exclusive four-pion production at $\sqrt{s_{\rm{NN}}} = 5.36 $ TeV measured using the ALICE detector in Run 3. The left-hand-side plot (a) shows the  production of $\rho$' and the right-hand-side plot (b) shows the first observation of  charmoniums ($J/\psi$ and $\psi$(2S)) in the four pion channel.}
    \label{fig:four_pi_run3}
\end{figure}

After the completion of the Long Shutdown 2, ALICE has a dedicated program on the physics of UPCs in Run 3 in nucleus-nucleus and proton-nucleus collisions~\cite{Citron:2018lsq,dEnterria:2025jgm}. ALICE has started collecting Pb--Pb collisions data since 2022. The upgrade has enabled the ALICE Collaboration to take data using the streaming readout mode with an interaction rate of up to 50 kHz. As a result, ALICE now operates as mainly triggerless detector at the online level. It opened an opportunity to study exotic resonances and to collect high statistics, which was not possible in Run 2. For example, the studies discussed in the previous sections, such as the azimuthal correlation, excited $\rho'$ studies, can now be performed with higher precision and reduced kinematical biases or systematic effects associated with triggers.
Additionally, we have observed several new processes in the ultra-peripheral collisions at the center-of-mass energy of $\sqrt{s_{\rm{NN}}}=5.36 $ TeV.  The invariant mass distribution of the exclusively photoproduced $\rho'$ in four-pion channel in UPCs is shown in Fig.~\ref{fig:rho_prime_run3}. Finally, for the first time, some new interesting channels have been observed in UPCs, including the production of $J/\psi$ mesons decaying into four-pion events Fig.~\ref{fig:jptofourpi_run3}, a di-kaon-di-pion channel, and a six-pion channel. 

\begin{figure}[!!hhh]
    \centering
    \begin{subfigure}{0.48\linewidth}
        \centering
        \includegraphics[width=\linewidth]{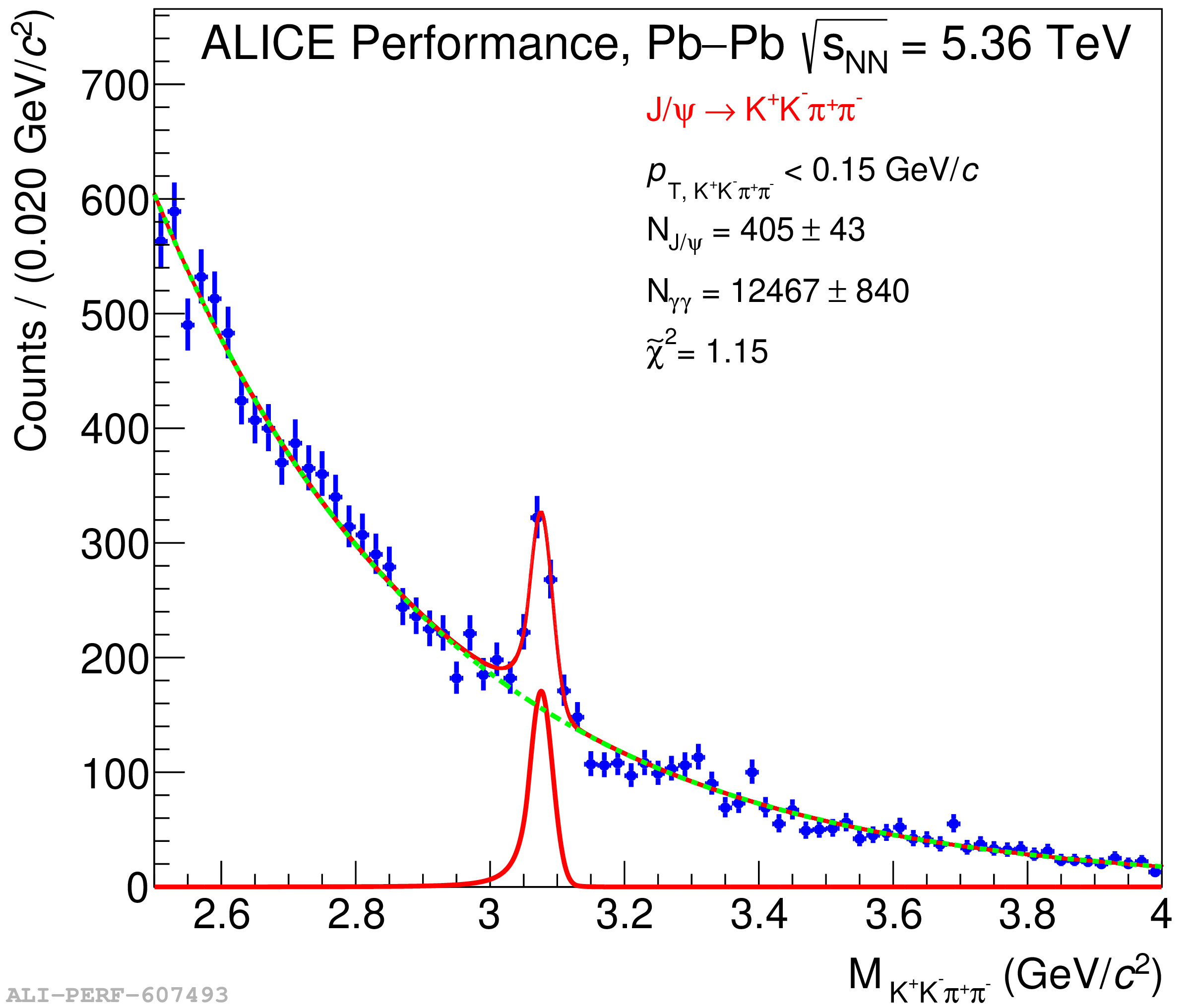}
        \label{fig:jp2pi2k}
    \end{subfigure}
    \hfill
    \begin{subfigure}{0.48\linewidth}
        \centering
        \includegraphics[width=\linewidth]{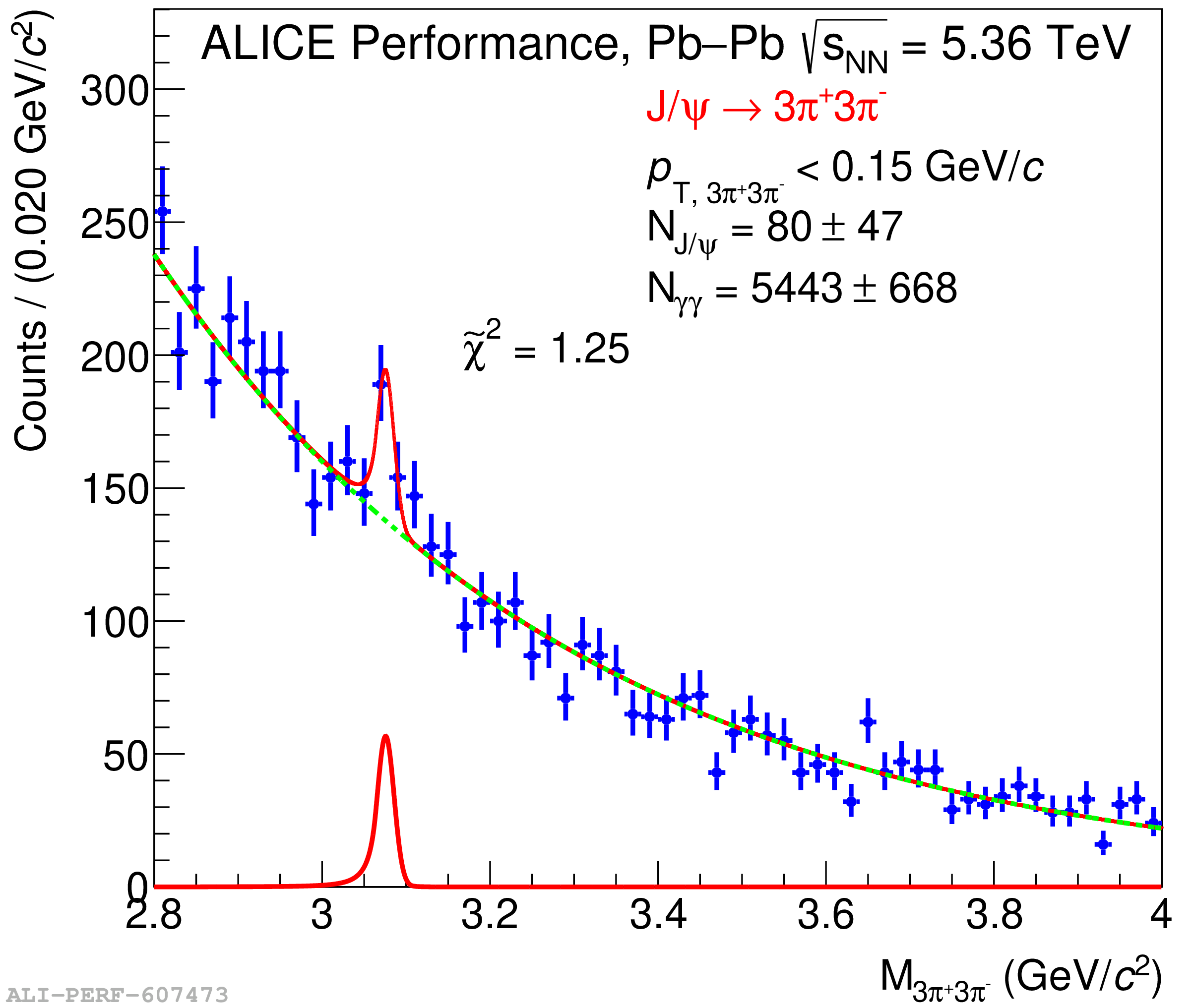}
        \label{fig:jpsixpi}
    \end{subfigure}
    \caption{Invariant mass of $J/\psi$ production in di-kaon and di-pion channel (left) and six pion production (right).}
    \label{fig:second_pair}
\end{figure}

\section*{Summary}
The ALICE Collaboration has an exciting program to study UPC processes. 
 The first measurement of impact parameter dependence of azimuthal correlations for exclusive $\rho^0$ production using Run 2 data was presented. This measurement showed that the magnitude of the modulation increases with increasing impact parameter. The first measurement of excited $\rho$ in the four pion channel in Run 2 at $\sqrt{s_{\rm{NN}}} = 5.36 $ TeV was discussed. The measurement favors the presence of two resonances $\rho$(1450)
and $\rho$(1700). Additionally, we presented performance studies for $\rho$' in the four-pion channel in Run 3, and presented the first observation of new decay channels of $J/\psi$ into four-pions, six-pions, and di-pion di-kaon. These new measurements in Run 3 will help to continue the study of highly dense gluonic matter in a more systematic and multi-differential way. 

\newpage
\section*{References}

\bibliography{citation}{}


\end{document}